\documentstyle[psfig,subfigure,lscape]{mn}

\begin{document}
\newcommand{\goodgap}{%
 \hspace{\subfigtopskip}%
 \hspace{\subfigbottomskip}}

\title[The Canada-UK Deep Submillimetre Survey VIII]
{The Canada-UK Deep Submillimetre Survey VIII: Source Identifications
in the 3-hour field}
\author[D. Clements et al.]{Dave Clements$^{1,2}$, 
Steve Eales$^1$\thanks{Comments or queries about the 
paper should be sent to Steve Eales 
at sae@astro.cf.ac.uk}, Kris Wojciechowski$^1$, Tracy Webb$^3$,
\newauthor
Simon Lilly$^4$, 
Loretta Dunne$^1$,
Rob Ivison$^5$,
Henry McCracken$^6$,
\newauthor
Min Yun$^7$, Ashley James$^1$, Mark Brodwin$^8$, Olivier Le F\`evre$^9$ and
Walter Gear$^1$\\ 
$^{1}$Department of Physics and 
Astronomy, Cardiff University, P.O. Box 913, Cardiff CF24 3YB, UK \\
$^{2}$ Imperial College, Blackett Laboratory, Prince Consort Road,
London SW7 2BZ\\ 
$^{3}$ Sterrewacht Leiden, Postbus 9513, 2300 RA Leiden,
the Netherlands\\
$^{4}$ Institut f\"ur Astronomie, ETH H\"onggerberg, HPF G4.1,
CH-8093, Z\"urich, Switzerland\\
$^{5}$ Astronomy Technology Centre, Royal Observatory, 
Blackford Hill, Edinburgh EH9 3HJ, UK\\
$^{6}$ University of Bologna, Department of Astronomy,
via Ranzani 1, 40127 Bologna, Italy\\
$^7$ Department of Physics and Astronomy, University of Massachusetts,
640 Lederle Graduate Research Center, Amherst, MA01003, USA\\
$^8$ Department of Astronomy, University of Toronto, 60 St. George
Street, Toronto, Ontario, Canada, M5S 3H8\\
$^9$ Labatoire d'Astrophysique de Marseiles, Traverse du Siphon,
13376 Marseille Cedex 12, France}

\maketitle                  

\begin{abstract}
We present optical, near-infrared and radio observations
of the 3-hour field of the Canada-UK Deep Submillimetre Survey.
Of the 27 submillimetre sources in the field,
nine have secure identifications with either a radio source
or a near-IR source. We show that the
percentage of sources with secure identifications
in the CUDSS
is consistent with that found for the bright `8 mJy' submillimetre
survey, once allowance is made for the different submillimetre
and radio flux limits. Of the 14 secure identifications
in the two CUDSS fields,
eight are VROs or EROs,
five have colours typical of normal galaxies, and one is a radio source
which has not yet been detected at optical/near-IR wavelengths.
Eleven of the identifications have optical/near-IR
structures which are either disturbed or have some peculiarity which
suggests that the host galaxy is part of an interacting system.
One difference between the CUDSS results and the results
from the 8-mJy survey is the large number of low-redshift objects
in the CUDSS; we give several arguments why these are
genuine low-redshift submillimetre sources rather than
being
gravitational lenses which are gravitationally amplifying a high-$z$ 
submillimetre
source.
We construct a 
$K-z$ diagram for various classes of high-redshift
galaxy and show that the SCUBA galaxies are on average
less luminous than
classical radio galaxies, but are very similar in both their
optical/IR luminosities and their colours to
the host galaxies of the radio sources detected in
$\mu$Jy radio surveys.
\end{abstract}


\section{Introduction}       
The luminous high-redshift dust sources discovered by the
SCUBA submillimetre and MAMBO millimetre surveys 
(Smail, Ivison \& Blain 1997; Hughes
et al. 1998; Barger et al. 1998;
Eales et al. 1999; Bertoldi et al.
2000,2001) are almost certainly 
of great significance for our understanding of
galaxy formation. 
The ultimate energy source in these objects
is hidden by dust but the two
obvious possibilities are that (1) the dust is being heated by a hidden active
nucleus or (2) the dust is being heated by a luminous population of stars.
The first of these can now largely be ruled out
because of the  
failure of the XMM/Newton and
Chandra telescopes
to detect strong X-ray emission from 
many of the dust sources (e.g. Ivison et al. 2002; Almaini et al. 2003;
Waskett et al. 2003a; Alexander et
al. 2003). Estimates of the star-formation rates necessary
to produce the dust luminosity can be as 
high as $\rm 6 \times 10^3\ M_{\odot}$ year$^{-1}$
(Smail et al. 2003), enough to produce the stellar population of a massive
galaxy in $\sim 10^8-10^9$ years. 
Many authors have concluded that these dust sources
are the ancestors of present-day elliptical galaxies, basing their
arguments on estimates of the star-formation rate in the population
as a whole (Smail, Ivison and Blain 1997; Hughes et al. 1998;
Blain et al. 1999),
on estimates of the contribution of the sources to the extragalactic background
radiation (Eales et al. 1999), and on comparisons of the space-density
of the SCUBA/MAMBO sources (henceforth SMS) 
with the space-density of ellipticals in the
universe today (Scott et al. 2002; 
Dunne, Eales and Edmunds 2003). 

In view of the probable significance of this population, it is of great
importance to determine the optical counterparts
to the SMSs and measure their redshifts. If the SMSs are the ancestors
of present-day elliptical galaxies, what do the properties of the SMSs
tell us about how elliptical galaxies form?
There are two rival theories of the birth of an elliptical.
In the older of these
(Eggen, Lynden-Bell
and Sandage 1962; Larson 1975), an elliptical forms when
an individual gas cloud in the early universe collapses,
most of the galaxy's stars forming during the 
collapse.
In the modern theory, the elliptical
forms as the result of a sequence of galaxy mergers.
This may occur over a relatively long period of cosmic
time,
with a burst of star formation being triggered
during each merger, or
it may occur as the result
of a few mergers at high redshift (Cole et al. 2000; Percival et al.
2003).
If an SMS does represent
one of these galaxy-building bursts of star formation, the redshifts
of the SMSs are clearly crucial for 
determining the correct model for elliptical formation.
If the older theory is correct, for example, the redshift
distribution of the SMSs should probably have a pronounced
peak, corresponding to
the epoch in which most ellipticals formed. 

Unfortunately one of the major problems in understanding
this population has been the difficulty of determing the optical
counterparts and measuring the redshifts of the SMSs.
The obstacles here are the large
errors on the positions of the SMSs, which often make
it difficult to determine
the optical/IR counterpart to the SMS,
and the faintness of these counterparts, which make it
difficult to measure a redshift.
Until recently, the recourse of most groups has been to
try to detect the SMSs at radio wavelengths,
since the surface density of radio sources in deep VLA radio surveys is
low enough that it is possible to be confident that apparent radio
counterparts to SMSs are not chance coincidences.
Once an SMS has been securely identified with a radio source, the
accurate radio position can be used to
determine 
the optical/IR counterpart. Furthermore, Carilli and Yun (1999)
pointed out that, if SMSs are star-forming galaxies like those in the
universe today, it is possible to estimate the redshift of
the SMS from the ratio of radio to submillimetre flux.
Fortunately, a significant fraction of the SMSs are also
faint radio sources.
Ivison et al. (2002), for example, found that 60\% of the SMSs
in the `8 mJy' SCUBA survey are also radio sources.
The optical objects found at the radio positions are usually faint,
often appear to be  
merging or interacting systems (Lilly et al. 1999; Ivison et
al. 2000; Webb et al. 2003a), and often have very red optical-infrared (I-K) colours,
with a significant number being as red as the `Extremely Red Objects'
(Ivison et al. 2002; Webb et al. 2003a).

Recently Chapman et al. (2003a) have taken a major step forward
by measuring the redshifts for a significant number of SMSs
with accurate radio positions.
Rather surprisingly, given the dust in these objects, this group
succeeded in detecting Lyman $\alpha$ and other $UV$ lines with the
Keck Telescope from 10 SMSs. The redshifts they have
measured lie in the range $\rm 0.8 < z < 4$, although
because of the requirement for accurate radio positions,
and because the ratio of radio to submillimetre flux is expected
to fall with redshift (Carilli and Yun 1999),
this distribution
may well be skewed towards low redshifts.
Nevertheless, the wide range of redshifts is in better agreement
with modern ideas about the formation of ellipticals than with the
older theory. The result that a large fraction of SMSs are merging
or interacting systems is also in better agreement with these ideas.

In this paper we describe the results of our attempts to determine
the optical and counterparts to the SMSs in the 3-hour field
of the Canada-UK Deep Submillimetre Survey.
Our cosmological assumptions in this paper are a
concordance universe with $\rm \Omega_{\Lambda}=0.7$ and
$\rm \Omega_M=0.3$ and a
Hubble constant of
$\rm 75\ km\ s^{-1}\ Mpc^{-1}$.

\section{The Survey}

The Canada-UK Deep Submillimetre Survey (Eales et al. 1999)
is
one of the largest of the deep SCUBA submillimetre surveys.
The basic survey consists of deep 850$\mu$m images of two
fields at a right ascension
of 3$^h$ and 14$^h$. Each field is about $\rm 6 \times 8$ arcmin$^2$
in size and the 3$\sigma$ sensitivity at 850$\mu$m is about 3 mJy.
This is the eighth paper describing the results from the
survey. The first two papers
(Eales et al. 1999; Lilly et al. 1999)
describe the submillimetre and optical results from
initial surveys of parts of the two fields, together with
the results from a survey of a third smaller field at a
right ascension of 10$^h$.
Paper III (Gear et al. 2000) describes millimetre
interferometry of the brightest source in the 14$^h$ field.
Paper IV (Eales et al. 2000) describes the submillimetre
observations of the 14$^h$ field.
Paper V (Webb et al. 2003b) describes an investigation
of the cross-clustering between the SCUBA sources and the
Lyman break galaxies in the two fields.
Paper VI (Webb et al. 2003c; henceforth W2003) describes the submillimetre
survey of the 3-hour field.
Paper VII (Webb et al. 2003a) describes the follow-up
optical/IR observations of the 14-hour field.
This paper describes the follow-up optical/IR
observations of the 3-hour field. A final paper (Eales et al.,
in preparation) will describe
an investigation of galaxy evolution in the submillimetre
waveband using the results from the survey.
The 27 sources in the 3-hour field (W2003) are listed in Table 1.

\begin{table*}
\begin{tabular}{|l|l|l|l||l}
\multicolumn{5}{|c|}{Table 1. Submillimetre Sources}\\
\hline
(1)&(2)&(3)& (4) & (5)\\
Name & RA (J2000) & Dec (J2000) & S/N & $\rm S_{850 \mu m}$/mJy \\
\hline
CUDSS 3.1 & 03 02 44.55 & 00 06 34.5 & 7.4 & 10.6$\pm$1.4 \\
CUDSS 3.2 & 03 02 42.80 & 00 08 1.50 & 6.7 & 4.8$\pm$0.7  \\
CUDSS 3.3 & 03 02 31.15 & 00 08 13.5 & 6.4 & 6.7$\pm$1.0  \\
CUDSS 3.4 & 03 02 44.40 & 00 06 55.0 & 6.2 & 8.0$\pm$1.3  \\
CUDSS 3.5 & 03 02 44.40 & 00 08 11.5 & 5.8 & 4.3$\pm$0.7  \\
CUDSS 3.6 & 03 02 36.10 & 00 08 17.5 & 5.4 & 3.4$\pm$0.6  \\
CUDSS 3.7 & 03 02 35.75 & 00 06 11.0 & 5.3 & 8.2$\pm$1.5  \\
CUDSS 3.8 & 03 02 26.55 & 00 06 19.0 & 5.0 & 7.9$\pm$1.6  \\
CUDSS 3.9 & 03 02 28.90 & 00 10 19.0 & 4.6 & 5.4$\pm$1.2 \\
CUDSS 3.10 & 03 02 52.50 & 00 08 57.5 & 4.5 & 4.9$\pm$1.1  \\
CUDSS 3.11 & 03 02 52.90 & 00 11 22.0 & 4.0 & 5.0$\pm$1.3 \\
CUDSS 3.12 & 03 02 38.70 & 00 10 26.0 & 4.0 & 4.8$\pm$1.2  \\
CUDSS 3.13 & 03 02 35.80 & 00 09 53.5 & 3.8 & 4.1$\pm$1.1  \\
CUDSS 3.14 & 03 02 25.78 & 00 09 7.50 & 3.5 & 5.1$\pm$1.5  \\
CUDSS 3.15 & 03 02 27.60 & 00 06 52.5 & 3.5 & 4.4$\pm$1.3  \\
CUDSS 3.16 & 03 02 35.90 & 00 08 45.0 & 3.4 & 2.8$\pm$0.8  \\
CUDSS 3.17 & 03 02 31.65 & 00 10 30.5 & 3.4 & 5.0$\pm$1.5 \\
CUDSS 3.18 & 03 02 33.15 & 00 10 19.5 & 3.3 & 3.9$\pm$1.2 \\
CUDSS 3.19 & 03 02 43.95 & 00 09 52.0 & 3.2 & 3.3$\pm$1.0 \\
CUDSS 3.20 & 03 02 53.30 & 00 09 40.0 & 3.2 & 3.4$\pm$1.1  \\
CUDSS 3.21 & 03 02 25.90 & 00 08 19.0 & 3.1 & 3.8$\pm$1.2  \\
CUDSS 3.22 & 03 02 38.40 & 00 06 19.5 & 3.1 & 3.1$\pm$1.0  \\
CUDSS 3.23 & 03 02 54.00 & 00 06 15.5 & 3.1 & 5.8$\pm$1.9 \\
CUDSS 3.24 & 03 02 56.80 & 00 08 8.00 & 3.0 & 5.1$\pm$1.7  \\
CUDSS 3.25 & 03 02 38.65 & 00 11 12.0 & 3.0 & 4.1$\pm$1.4  \\
CUDSS 3.26 & 03 02 35.10 & 00 09 12.5 & 3.0 & 3.6$\pm$1.2 \\
CUDSS 3.27 & 03 02 28.56 & 00 06 37.5 & 3.0 & 4.0$\pm$1.3 \\

\hline
\end{tabular}
\flushleft
(1) Source name. (2) \& (3) Position (RA and Dec) in J2000 coordinates
(4) Signal-to-noise with which the submillimetre source was detected (W2003),
(5) Flux at 850$\mu$m of source in mJy.

\end{table*}

\section{The Observations}

This field was originally observed in the optical and infrared wavebands
as part of the Canada-France Redshift Survey (Lilly et al. 1995).
It has been observed in the mid-infrared waveband
with the Infrared Space Observatory
(W2003; Flores et al. 2003) and in the X-ray waveband with XMM/Newton
(Waskett et al. 2003a). To determine the counterparts to the SCUBA sources,
we have used a new radio image obtained with the VLA, deep infrared
observations made with the UK Infrared Telescope (UKIRT) and with
the Canada-France Hawaii Telescope (CFHT), and optical observations
made with the CFHT and the Hubble Space Telescope.

\subsection{Radio Observations}

We observed this field at 1.4 GHz with the VLA in both the A and B
configuration. The reduced radio image and the source catalogue will
be presented elsewhere. The noise on the final image was 11 $\mu$Jy.

\subsection{Infrared Observations}

We obtained two complementary datasets for the field: observations
in the K-band with the infrared camera (UFTI) on the UKIRT and observations
in the K$'$-band with the infrared camera (CFHTIR) on the CFHT.

\subsubsection{UKIRT Observations}

The UFTI camera on the UKIRT uses a tip-tilt-correcting secondary mirror
to deliver images with high angular resolution to an array with small
pixels (0.0906 arcsec). It can thus provide very deep high resolution images,
which are useful not only for identifying the CUDSS sources but also
for providing morphological information which may help us to understand
their origin and nature. The small pixel size of the UFTI, however, 
has the drawback that many fields would need to be observed to
cover completely the region of the submillimetre survey.
At present, we have obtained UFTI images of 18 of the 27 sources.

We observed the 3-hour field
on the nights 9, 10, 12, 18 and 23 January 2000 and 20-22 October 2001
in the
K-band. The camera has
a 1024$^2$ HgCdTe focal-plane array which gives
a field-of-view of $93 \times 93$ arcsec$^2$. Our observations consisted
typically of a series of nine short (120s in the first run, 80s in the
second) integrations, made in a semi-random pattern over a $9 \times
9$ arcsec$^2$ area of sky. After the first group of nine observations,
the telescope was offset by 1 arcsec, and the same pattern was repeated.
Each of these cycles consists of 18 minutes of integration time (12 minutes
for the later run). We carried out between six and 10 of these cycles for
each target. To calibrate our observations, 
we observed stars from the
list of UKIRT faint photometric standards several times
each night, principally
FS7, FS11 and FS30.

We carried out the reduction of the data from each nine-integration
cycle using the ORACDR pipeline
system, which contains procedures for sky subtraction, flat fielding,
the removal of bad pixels, and the coaddition and registration of 
the individual observations. The output of the pipeline is a fully-reduced
image of the data from each cycle.
For each target, we then aligned and added the images produced by the
pipeline, using routines from the STARLINK CCDPACK library.
The faintest objects visible on the final images have magnitudes
between $\rm K \sim 20.5$ and $\rm K \sim 21.5$.

\subsubsection{CFHT Observations}

We observed the 3-hour field in the K$'$ band using the new near-infrared
camera (CFHTIR) on the CFHT during the period 9-15 Jan 2001. The seeing
was typically 0.8-1.0 arcsec. CFHTIR has $\rm 1024 \times 1024$
pixels, each with a size of 0.211 arcsec, giving a field-of-view of
$\rm 3.6 \times 3.6$ arcmin. We covered two thirds of the area of the
3-hour submillimetre survey with a mosaic of 30-second exposures.
We reduced the data using IRAF routines (see Webb et al. 2003a for
more information), producing a single image covering 23 of the 27
CUDSS sources.
The total integration time at a typical point
in the image is 2.7 hours and the faintest visible objects
have $\rm K \sim 21.6$.

\subsubsection{Deeper Images}

As a result of these observations, there was often more than
one image of a CUDSS source. In order to obtain as deep an image
as possible, we coadded the images. We did this using standard
procedures within the STARLINK library. We first extracted the
relevant section of the large CFHT image, binned the UFTI image
so that it had pixels of the same size as the CFHT image, aligned
the images using objects visible on both images, and scaled
the images onto a common photometric scale. 
We then measured the noise on each image, and then added the images
using as weights the inverse square of the measured noise.
We astrometrically calibrated the final images using objects
which were visible in both the K-band images and in the Canada-France
Deep Field I-band image (see below). The good agreement between
the radio and K-band positions (\S 4) implies that in most cases
the accuracy of the K-band positions is better than 0.5 arcsec.

The final images are shown in Figure 1. We used the SEXTRACTOR image-detection
package to produce catalogues of sources for use in our identification
analysis. We obtained K-band magnitudes of each potential identification
using a circular aperture with a diameter of 3 arcsec. 

\subsection{Optical Observations}

The three-hour field was observed in the optical waveband as part of
the Canada-France Deep Fields Survey (CFDF, McCracken et al. 2001).
The images obtained as part of this survey consisted of U, B, V and
I-band images, each covering an area of 0.25 degrees$^2$, and reaching
a 3$\sigma$ limiting AB magnitude of 26.98, 26.38, 26.40, and
25.62, respectively. We used the CFDF images to obtain optical magnitudes
for the potential identifications found on the K-band image. In all cases,
we used the same 3-arcsec aperture as we had used to measure the K-band
magnitudes. The CFDF I-band image of each CUDSS source is shown alongside
the K-band image in Figure 1.

For a few of the CUDSS sources there are 
images taken with the Hubble Space Telescope (HST).
Brinchman et al. (1998) obtained three images
with the Wide Field and Planetary Camera 2 (WFPC2) in which CUDSS
sources fall. These images were taken through the F814W filter
and had an integration time of 6700s. 
We have also obtained a few WFPC2 images specifically to follow up
the CUDSS sources. These
images
were also taken
through the F814W filter and had an integration time of 7000s. 
The HST data that exists for the CUDSS fields is described
in more detail in Webb et al. (2003a).

Note that optical and IR magnitudes given in this paper are based
on the Vega zeropoints unless otherwise stated.

\section{Identification Procedure}

The biggest problem in 
determining the optical counterparts to the SCUBA sources
are the large errors in the positions of the sources.
The size of the errors is poorly known because of the
uncertain effect on the positions of nearby faint sources
which are too faint to detected
individually. Various authors
have tried to model this effect (Eales
et al. 2000; Hogg 2001; Scott et al. 2002). Eales et al. (2000), for
example, carried out
an investigation of the positional errors in the CUDSS, using two
different methods. They added artificial sources to the real
SCUBA images and then compared the positions determined by the
source-detection algorithm to the true positions. They also
carried out a full-scale Monte-Carlo simulation of one
of the CUDSS fields and compared the input and output positions
of the sources.
They
concluded that
between 10 and 20\% of CUDSS sources have measured positions which differ
from the true position by $>$6 arcsec. 
Scott et al. (2002) have also examined the effect of adding artificial
sources onto their real SCUBA maps. They concluded that the
mean positional error in their `8 mJy survey' 
is $\simeq$3-4 arcsec. 
Since the size of the errors is poorly known, it is impossible to
use Bayesian statistical techniques (e.g. Sutherland and Saunders
1990). Instead, we and others have adopted the frequentist 
technique of looking for objects close to the SCUBA position
and then estimating the probability of that object being a chance
coincidence (e.g Lilly 1999; Ivison et al. 2002; Serjeant et al.
2003). 

The first step in the procedure was to select a radius 
within which to look for possible counterparts to the SCUBA source.
We chose a radius of 8 arcsec, for the practical reason that
at larger radii we cannot distinguish a genuine association
from a chance coincidence. It is possible, of course, that placing
this limit on the search radius will have resulted in our missing some
genuine associations. Our investigation of the positional errors
in the CUDSS (see above) implies that we will have missed
$\simeq$5-8\% of the associations. However, this estimate is
based on simulations. We will show later that we can 
now empirically estimate the true distribution of positional
errors for SCUBA sources
(\S 6). This empirical investigation implies that our
earlier estimates of the positional errors for SCUBA sources
were too pessimistic. 

The most useful image for our identification analysis is the
radio image because the surface-density of radio sources is
sufficiently low, even
at the $\mu$Jy level, that it is possible to determine 
whether a radio source is genuinely associated with a SCUBA
source with high statistical certainty.
As the first step in analysis, we looked for sources
within 8 arcsec of the SCUBA position 
with a peak flux brighter
than 40$\mu$Jy (3.6 $\sigma$). 
There were 11 sources brighter than this limit within 8 arcsec
of the 27 CUDSS sources. Any real source should have an angular
size at least as large as the angular resolution of the VLA
at this frequency (FWHM of 1.4 arcsec), and after analysing the
source structures with the AIPS program JMFIT, we eliminated
two sources which had structures inconsistent with the
VLA beam. The probability of a source detected at
$>$3.6$\sigma$ within the search area
being the result of noise
is $\simeq$0.02. Since we have searched around
27 CUDSS sources, the expected number of false radio sources
is $\simeq$0.54. Therefore, it is possible that one of
the nine radio sources is spurious. However, seven of the
nine radio sources are coincident with galaxies (see below),
and so are definitely genuine. The exceptions are the sources
associated with 3.17 and 3.27.

Given that a radio source is genuine, the probability
of it not being associated with the SCUBA source is
\smallskip
$$
p = 1 - exp(-d^2 \pi n) 
$$
\smallskip
in which $d$ is the offset between the SCUBA source and
the radio position, and $n$ is the surface density of radio
sources. We calculated the surface density of radio
sources brighter than 40$\mu$Jy using the source counts
from the Hubble Deep Field (Richards et al. 2000).
The probabilities and offsets are listed in Table 2.
Seven of the radio sources have probabilities of being
chance coincidences of $<$1\%. The remaining two have probabilities
of being chance coinicdences of 3 and 4\%. Therefore, all of these
radio sources are almost certainly associated with the nine CUDSS
sources. 

The surface density of objects on the infrared images in Figure 1
is much higher and so we have to use a more sophisticated
technique for discriminating between chance coincidences
and genuine associations.
Since common sense says that a 17th magnitude galaxy two arcsec
from the SCUBA position is less likely to be a chance coincidence
than a 24th magnitude galaxy 
(because 17th magnitude galaxies are much rarer than 24th magnitude
galaxies),
we need to find a statistic which incorporates
the magnitude of the possible association.
We have used the statistic suggested by Downes et al. (1986)
to calculate the probability that a candidate galaxy on an infrared image
within 8 arcsec of the
SCUBA position is actually unrelated to the SCUBA source:
\smallskip
$$
S = 1 - exp(-d^2 \pi n(<m))
$$
\smallskip
in which $n(<m)$ is now the surface density of galaxies brighter
than the magnitude ($m$) of the possible association. 

The expression above looks like a probability, but it is not because it
does not take account of the galaxies on the image which are
fainter than the magnitude of the candidate galaxy. 
If one of these galaxies had been closer to the SCUBA position,
it might have had a lower value of $S$, and therefore in deriving the
sampling distribution for $S$ this possibility has to be taken into
account.
Downes et al. (1986) describe an analytic technique for
determining the sampling distribution of $S$.
However, because of the effect of clustering 
and because images do not always have a uniform depth, it is
preferable to use a Monte-Carlo simulation to determine the
probability that a candidate identification which is actually
physically unrelated to
the SCUBA source has a value of $S$ as low as the measured value.
We
calculated $S$ 
for each object within 8 arcsec of the SCUBA position, and then
used the Monte-Carlo technique described by
Lilly et al. (1999) to estimate the probability ($P$) that 
a physically unrelated object would have such a low value
of $S$.
We have listed
in Table 2 all the objects which have values for this probability $<0.3$.
As in our earlier paper, we found that the value of $P$ was typically
between six and seven times the value of $S$.

We used the infrared images in preference to the optical images
for this analysis, because SCUBA galaxies are generally quite
red 
(Smail et al. 2000;
Ivison et al. 2002), and so the infrared images make it possible to
discriminate between genuine associations and coincidences
with greater statistical precision than is possible
with optical images.
There are, however, two SCUBA sources for which 
there is no object close to the SCUBA position visible on the 
infrared image but for which there is an object visible on the CFDF I-band
image. In the case of these two sources, we applied our analysis to
the I-band data, although for these sources we calculated $P$ using
the analytic relationships in Downes et al. (1986), rather than
applying the full Monte-Carlo analysis.

Almost all the CUDSS sources for which there are objects 
on the infrared images with
values of $P$ less than 0.1 also have radio associations.
In most cases, the radio sources coincide, to within the positional
errors, with the infrared sources. There are only two CUDSS
sources which do not have radio associations but which have possible
infrared associations. One, CUDSS 3.2, has a value for $P$ of
0.02. The second, CUDSS 3.5, has a value for $P$ of
0.08. If one was considering CUDSS 3.5 in isolation, one would conclude
that the galaxy is genuinely associated with the SCUBA source, since
the probability of it being a chance projection is only 8\%.
However, it is not possible to consider the source entirely
in isolation. The 3-hour catalogue contains 27 SCUBA sources and,
{\it even if there are no galaxies associated with these SCUBA sources,
one expects to find $0.1 \times 27\simeq 3$ objects on 
the infrared images with values
of $P\simeq 0.1$.} For this reason, we have decided not to classify
this galaxy as a secure identification. In Table 2, we have divided possible
identifications into two classes. 
We have classified all but one of the SCUBA sources
with close radio sources, as well as CUDSS 3.2, as having secure 
identifications.
The source we have omitted is CUDSS 3.27, which has two close radio
sources. One of these is almost certainly the correct association, but
as we are not sure which, we have omitted it from the secure
class.
We can make a rough estimate of the probability that one of these
proposed secure identifications is actually wrong by adding the values of $P$
in Table 2. The total is 0.11, which means the chance of one of these
nine
secure identifications being wrong is about 10\%.
Our second class of identifications are suspected identifications. We
have placed CUDSS 3.5 in this class.
Apart from CUDSS 3.27,
we have also placed two other SCUBA sources in this class. 
For these sources, the statistical evidence 
that the proposed identification is correct is rather weak,
but there is circumstantial evidence,
based on the similarity of the colour or structure of the
galaxy to known SCUBA galaxies, that the identification
is correct (see notes on sources).

\begin{table*}
\begin{tabular}{|l|l|l|l||l|l|l|l}
\multicolumn{7}{|c|}{Table 2. Identifications}\\
\hline
(1)&(2)&(3)& (4) & (5) & (6) & (7) & (8)\\
Name & RA (J2000) & Dec (J2000) & Radio, K or & Flux or & Distance & $P$ & Status \\
 &                &             & I-band &  magnitude & from SCUBA &  & \\
 &                &             &        &            & position & & \\
\hline
3.1     & 03 02 44.84 & 00 06 32.0 & K & 19.06$\pm$0.02 & 4.7 & 0.27  & ...\\
\hline
{\bf 3.2} & {\bf 03 02 42.80} & {\bf 00 08 02.5} & {\bf K} & {\bf 18.64$\pm$0.02} & {\bf 1.1} & {\bf 0.021} & {\bf secure}  \\
\hline
3.3 & ...... & ...... & .. & ..... & ... & .... & ... \\
\hline
3.4 & 03 02 44.59 & 00 06 54.9 & K & 19.64$\pm$0.03 & 2.8 & 0.21 & suspect \\
\hline
3.5 &  03 02 44.45 &  00 08 11.1 &  K &  21.4$\pm$0.2 &  0.8 & 0.08 & suspect \\ 
\hline
{\bf 3.6} & {\bf 03 02 36.14} & {\bf 00 08 16.8} & {\bf R} & {\bf 43$\pm$12$\mu$Jy} & {\bf 1.0} & {\bf 0.0006 } & {\bf secure} \\
          & {\bf 03 02 36.14} & {\bf 00 08 16.9} & {\bf K} & {\bf 21.48$\pm$0.21}  & {\bf 0.9} & {\bf  0.066 } & ... \\
\hline
{\bf 3.7} & {\bf 03 02 35.89} & {\bf 00 06 11.5} & {\bf R} & {\bf 44$\pm$12$\mu$Jy}& {\bf 2.2} & {\bf 0.0029 } & {\bf secure} \\
          & {\bf 03 02 35.90} & {\bf 00 06 12.0} & {\bf K} & {\bf 20.45$\pm$0.09}  & {\bf 2.4} & {\bf 0.19   } & ... \\
         &  03 02 35.70 & 00 06 09.5 & K & 20.54$\pm$0.10   & 1.6 & 0.11   & ...  \\
\hline
{\bf 3.8} & {\bf 03 02 26.15} & {\bf 00 06 24.1} & {\bf R} & {\bf 683$\pm$21$\mu$Jy} & {\bf 7.8} & {\bf 0.038  } & {\bf secure} \\
          & {\bf 03 02 26.16} & {\bf 00 06 24.2} & {\bf K} & {\bf 14.56$\pm$0.003 }  & {\bf 7.8} & {\bf 0.074 } & ... \\
\hline
{\bf 3.9} & {\bf 03 02 28.95} & {\bf 00 10 18.6} & {\bf I} & {\bf 24.23$\pm$0.05}   & {\bf 0.9} & {\bf 0.12 } & ... \\
\hline
{\bf 3.10}&  {\bf 03 02 52.50} & {\bf 00 08 56.4} & {\bf R} & {\bf 154$\pm$34$\mu$Jy}& {\bf 1.1} & {\bf 0.00076} & {\bf secure} \\
          &  {\bf 03 02 52.50} & {\bf 00 08 56.4} & {\bf K} & {\bf 16.31$\pm$0.005}  & {\bf 1.1} & {\bf 0.0035 } & ... \\
\hline
3.11     & 03 02 52.85       & 00 11 22.1     & I & 25.0$\pm$0.4            & 0.8 & 0.15     & ...  \\
\hline
3.12     & ......       & ......     & ..& .....            & ... & ....    & ...  \\
\hline
3.13     &  03 02 36.06 & 00 09 58.3 & K & 17.33$\pm$0.01   & 6.1 & 0.11   & ...  \\
\hline
3.14     &  03 02 25.68 & 00 09 06.2 & K & 20.64$\pm$0.23   & 2.0 & 0.13    & suspect \\
\hline
{\bf 3.15}&  {\bf 03 02 27.73} & {\bf 00 06 53.5} & {\bf R} & {\bf 226$\pm$12$\mu$Jy}& {\bf 2.2} & {\bf 0.0029 } & {\bf secure} \\
          &  {\bf 03 02 27.72} & {\bf 00 06 53.2} & {\bf K} & {\bf 18.33$\pm$0.02}   & {\bf 2.0} & {\bf 0.076  } & ... \\
\hline
3.16     & ......       & ......     & ..& .....            & ... & ....    & ...  \\
\hline
{\bf 3.17}&  {\bf 03 02 31.80} & {\bf 00 10 31.2} & {\bf R} & {\bf 44$\pm$12$\mu$Jy} & {\bf 2.3} & {\bf 0.0033 } & {\bf secure} \\
          &  03 02 31.52 & 00 10 28.7 &  K & 18.52$\pm$0.03   & 2.7 & 0.068  & ...  \\
\hline
3.18     & ......       & ......     & ..& .....            & ... & ....  & ...   \\
\hline
3.19     & ......       & ......     & ..& .....            & ... & .... & ...    \\
\hline
3.20     & ......       & ......     & ..& .....            & ... & .... & ...    \\
\hline
3.21     & ...... & ...... &  ..& ..... &     ... & ....   & ...  \\
\hline
3.22     & ......       & ......     & ..& .....            & ... & .... & ...     \\
\hline
3.23     &  03 02 54.06 & 00 06 18.1 & K & 20.75$\pm$0.26   & 2.8 & 0.20 & ...    \\
\hline
{\bf 3.24}&  {\bf 03 02 56.58} & {\bf 00 08 06.6} & {\bf R} & {\bf 122$\pm$32$\mu$Jy}& {\bf 3.6} & {\bf 0.0082 } & {\bf secure} \\
          &  {\bf 03 02 56.57} & {\bf 00 08 06.5} & {\bf K} & {\bf 19.26$\pm$0.07}   & {\bf 2.8} & {\bf 0.16   } & ... \\
\hline
{\bf 3.25}&  {\bf 03 02 38.59} & {\bf 00 11 05.3} & {\bf R} & {\bf 353$\pm$12$\mu$Jy}& {\bf 6.8} & {\bf 0.028  } & {\bf secure} \\
          &  {\bf 03 02 38.58} & {\bf 00 11 05.5} & {\bf K} & {\bf 20.56$\pm$0.17}   & {\bf 6.5} & {\bf 0.64   } & ... \\
\hline
3.26     &  03 02 34.92 & 00 09 10.7 & K & 19.58$\pm$0.09   & 3.2 & 0.14   & ...  \\
\hline
3.27     &  03 02 28.53 & 00 06 45.0 & R & 43$\pm$12$\mu$Jy & 7.5 & 0.035   & ... \\
         &  {\bf 03 02 28.67} & {\bf 00 06 41.2} & {\bf R} & {\bf 49$\pm$12$\mu$Jy} & {\bf 3.9} & {\bf 0.0097 } & {\bf suspect}\\
         &  03 02 28.50 & 00 06 43.2 & K & 20.27$\pm$0.08   & 5.7 & 0.62  & ...  \\
\hline
\end{tabular}
\flushleft
(1) Source name. (2) \& (3) Position in J2000 coordinates
of the possible counterpart. (4) The waveband in which the possible
counterpart was found. An R indicates the counterpart was found on
our 1.4 GHz radio image, an I or a K indicate the standard optical/infrared
bands. (5) The flux density in $\mu$Jy of the counterpart if
it was found on the radio image; otherwise the I or K-band magnitudes
of the counterpart. The errors on the I- and K-band magnitudes
do not include the calibration error, which
is about 0.05 mags. (6) The distance in arcsec between the
position of the possible counterpart and the submillimetre position.
(7) The probability that the counterpart is not physically associated with
the SCUBA source. (8) Our assessment of the proposed identification based
on the criteria described in \S4.

\end{table*}

\begin{figure}

\vspace{8.0cm}

\caption{Optical (I-band) and near-infrared (K-band) 
images of the fields. The I-band image is from the
Canada-France Deep Field Survey. Each image
has a size of $\rm 20 \times 20$ arcsec$^2$, except for the images
of CUDSS 3.8, which have a size of $\rm 40 \times 40\ arcsec^2$. The circle
on each image is
centred on the SCUBA position and has a radius of 8 arcsec. 
The near-infrared image has the radio contours superimposed.
The five lowest contours are at intervals of
1$\sigma$, 2$\sigma$, 3$\sigma$, 4$\sigma$, 5$\sigma$
(11, 22, 33, 44 and 55 $\mu$Jy), with higher contours being at intervals
of 10$\sigma$, 20$\sigma$, 40$\sigma$ etc. 
}
\end{figure}

\addtocounter{figure}{-1}

\begin{figure}

\vspace{8.0cm}
\caption{---{\it continued}
}

\end{figure}

\begin{figure}

\vspace{8.0cm}
\caption{HST I-band images of five of the fields. Each image has a size
of $\rm 5 \time 5\ arcsec^2$, except for the image of CUDSS 3.8, which
has a size of $\rm 10 \time 10\ arcsec^2$
}

\end{figure}

\section{Notes on Individual Sources}

{\bf CUDSS 3.1:} 
This is the brightest source in either of the two CUDSS fields.
The object listed in Table 1 has a spectroscopic
redshift of 0.1952 (Hammer et al. 1995). There is a tentative 450$\mu$m
detection (W2003), but the 450$\mu$m position
is further from the position of the galaxy than
the 850$\mu$m position. 
Given the large SCUBA positional errors, it is possible that this
galaxy is the counterpart to the SCUBA source, but the large value
for $P$ means that we have no statistical evidence in favour of
this possibility.

\smallskip
\noindent {\bf CUDSS 3.2:}
This is the one source for which there is some circumstantial evidence
that gravitational lensing is important. The redshift estimated for 
the optical counterpart from the broad-band colours (\S 9.2) is 0.62,
whereas the estimated lower limit to the redshift of the SMS from the lack
of a radio detection (\S 9.1) is 1.7. The large difference in the redshifts
suggests that the SMS is behind the galaxy, with the submillimetre
flux being gravitationally amplified by the galaxy (Chapman et al. 2002).
The undisturbed morphology of the galaxy (Figure 1) is in agreement with
this hypothesis.

\smallskip
\noindent {\bf CUDSS 3.4:} We have classified the galaxy listed in Table 2 as
a suspected identification 
despite the lack of strong statistical evidence from its position and
magnitude, for the following reasons. First, the object has a very red colour
($I - K = 4.14$), which qualifies it as an Extremely Red Object (ERO), and
SCUBA sources are frequently found to be associated with EROs
(Ivison et al. 2000). Second,
there is a second much bluer object which is hardly visible on the
K-band image but is very prominent on the I-band image (Figure 1).
This is actually closer to the SCUBA position (2 arcsec) and has a slightly
lower value of $P$ (0.16, calculated from the statistics of the I-band
image). There are some faint signs on the CFDF I-band image (Figure 1),
although not on the HST image (Figure 2), of an interaction between the
two galaxies, which is also a common feature of SCUBA galaxies.

\smallskip

\noindent {\bf CUDSS 3.5:} There are some signs on both the K-band and I-band
images that this galaxy has a disturbed morphology. This is a common
feature of SCUBA galaxies, and the disturbed morphology adds some circumstantial
evidence to the statistical evidence that this is the correct identification.
We have classed this galaxy as a suspected identification for the reasons
described in \S 4.

\smallskip

\noindent {\bf CUDSS 3.6:} This is a secure identification, because the position
of the SCUBA source is only 1.0 arcsec away from a radio source.
The faint object visible on the K-band image (Figure 1)
at the radio position has a structure which looks like that of
an interacting galaxy.

\smallskip

\noindent {\bf CUDSS 3.7:} On the K-band
image there is a distinctive trapezium of sources. The SCUBA source
is detected at radio wavelengths, and the radio source is coincident with
the northern of the K-band sources. The northern and southern
K-band sources are
not detected at all in the CFDF I-band image, and the limits on their $I-K$
colours ($\rm > 4.4$ and $\rm > 4.1$) place them in the category
of
EROs. 
The eastern K-band source is just visible on the I-band image as
the northern of a pair of faint objects.
The $I-K$ colour of this source is
3.7, not as red as an ERO but falling within 
the class of VROs (Very Red Objects) 
according
to the definition of Ivison et al. (2002).
The western source is just barely detected in the I-band.
The $I-K$ colour is 3.6, making it a VRO.
The distinctive arrangement of the sources on the K-band image looks
remarkably like a case of gravitational lensing, but the
slightly different colours of the sources, and the fact that only
one is detected at radio wavelengths,
suggests that these sources are not four gravitational images. 
It therefore seems more likely
that the trapezium is actually a cluster of extremely-red
high-redshift galaxies.

\smallskip
\noindent {\bf CUDSS 3.8:} Despite the large offset between the radio position and
the SCUBA position, there is only a 4\% chance that this is a chance coincidence.
The peculiar morphology of this galaxy and the fact that it is a strong 15$\mu$m
ISO source (W2003) are compelling evidence that is this the correct identification
(\S 7). The morphology of the system is shown best in the HST image (Figure 2).
There are four galaxies and two point sources (presumably stars) visible. Three
of the galaxies are spirals. The fourth galaxy has very low surface brightness
and is just visible on the western edge of the HST image. The HST image shows
that the two brightest galaxies are interacting. The radio image (Figure 1) shows
that both of these galaxies are also radio sources. The radio emission is probably the
result of starbursts triggered in both galaxies by the interaction. 
The brightest galaxy has a spectroscopic redshift of 0.088.

\smallskip 

\noindent {\bf CUDSS 3.9:} There is nothing visible on the K-band image (Figure 1).
There is, however, a faint galaxy visible both on the I-band image from the
CFDF survey and on the HST image. The value of $P$ given in Table 2 has been calculated
from the statistics of the CFDF I-band image.

\smallskip

\noindent {\bf CUDSS 3.10:} This is a secure identification, with the radio position
only 1.1 arcsec away from the SCUBA position. We showed in our previous paper
(W2003) that this SCUBA source is also identified with an ISO 15$\mu$m source, the
ISO position also being only 1.5 arcsec from the SCUBA position. There is a bright
galaxy coincident with the 
radio position with a spectroscopic redshift of 0.176 (Hammer et
al. 1995). The I-band CFDF image (Figure 1) and especially
the HST image (Figure 2) suggest that the galaxy is involved in a merger.

\smallskip
\noindent {\bf CUDSS 3.11:} There is a very faint object visible on the CFDF I-band image
(Fig. 1). It is only 0.75 arcsec from the SCUBA position. However, because of the
high surface density of objects at this faint magnitude, the probability that it
is physically unrelated to the SCUBA source is 15\%.

\smallskip

\noindent {\bf CUDSS 3.14:} The probability that the object listed in Table 2, which
is 2 arcsec from the SCUBA position, is physically unrelated to
the SCUBA source
is 13\% and therefore above our threshold for a secure identification.
We have, however, listed it as a suspected identification because of the
other faint objects visible in the
CFDF I-band image, some
of which are even closer to the SCUBA position. The objects 
look remarkably like a
high-redshift cluster, with the object listed in Table 2
being the brightest galaxy in the cluster. 
This is circumstantial evidence in favour of our proposed identification
because, as we will show later in this paper (\S 10),
SCUBA galaxies can have optical/near-IR luminosities
as high as radio galaxies
or first-ranked cluster galaxies.

\smallskip

\noindent {\bf CUDSS 3.15:} This is a secure identification, with 
a radio source
lying only 2 arcsec from the SCUBA position. 
The source was also detected with ISO at 15$\mu$m (W2003).
On the K-band and CFDF I-band images (Fig. 1) the galaxy looks unexceptional,
but the HST image (Fig. 2) shows a ring
at the centre of the galaxy, about one arcsec across, 
encircling a point source. There
is also a faint arc on the HST image (labelled a in Figure
2). 
We cannot decide between two possible interpretations of this
system. One possibility is that the object is an example of a 
`collisional ring galaxy'. These objects are thought to be due
to the head-on collision between two galaxies, one of which
has travelled along the spin-axis of the other, striking the
disk of the second galaxy close to its centre (Appleton and Marston 1997).
From the multi-band optical and infrared photometry
of the identification we estimate that its redshift is $\simeq$0.7 (\S 9.2).
At this redshift, the physical size of
the ring would be typical of those seen in ring galaxies (Appleton and
Marston). In this interpretation, the arc seen on the HST
image would represent tidal debris from the collision. 
The alternative
interpretation is that the ring and arc represent a gravitational-lensing
phenomenom. The size of the ring is approximately what one expect
for an Einstein ring produced by a lens with the mass of a typical
galaxy. In view of the 15$\mu$m emission from this galaxy (\S 7),
we suspect the former explanation is the correct one.

\smallskip
\noindent {\bf CUDSS 3.17:} This is a secure identification, with the
radio source only 2.3 arcsec from the radio position. However, there
is nothing visible on either the I-band or K-band image at the
position of the radio source. 

\smallskip

\noindent {\bf CUDSS 3.22:} In W2003 we argued that this SCUBA source is identified
with a 15$\mu$m ISO source. The ISO position is 7.5 arcsec from the SCUBA position,
and so is indeed within our search radius. However, the two galaxies which are
the possible counterparts to the ISO source are both
outside the search radius. Indeed, it now seems likely
that it is the galaxy which is the furthest from the SCUBA position
which is the true counterpart to the ISO source, because this galaxy is
also a radio source. We therefore no longer think it is likely that the
ISO and SCUBA sources are related.
There are a number of possible identifications visible on the CFDF I-band image,
but none with a very low value of $P$.

\smallskip
\noindent {\bf CUDSS 3.24:} 
This is a secure identification, with a radio source only 3.6 arcsec from the
SCUBA position. The galaxy visible at the radio position (Fig. 1) is also
a 15$\mu$m ISO source (W2003). The $I-K$ colour of the galaxy is 3.50,
which puts in the category of VROs (Ivison et al. 2002).

\smallskip

\noindent {\bf CUDSS 3.25:} This is a secure identification. The radio source
is 6.8 arcsec from the SCUBA position but the probability that this is
a chance coincidence is only 3\%. The CFDF I-band image shows something which looks
like two interacting galaxies, which is circumstantial evidence that the
identification is correct.

\smallskip

\noindent {\bf CUDSS 3.26:} The probability of the galaxy being
unrelated to the SCUBA source
is 14\%, and thus the statistical evidence that this
is the correct identification is weak. A piece of circumstantial evidence
that this is the correct identifcation are the different morphologies visible
on the I-band and K-band images. The structure on the K-band image
extends to the south, suggesting that there may be two objects, a normal galaxy and
a very red object which only becomes visible on the K-band image. This is a situation which
has been seen for other SCUBA sources (Smail et al. 2002; Chapman et al. 2002).

\smallskip

\noindent {\bf CUDSS 3.27:} This is a rather peculiar field because there
are two radio sources, both of which have low values of $P$,
and it is not clear which is the correct identification. 
We have selected the radio source which is closest to
the SCUBA position (and thus has the lowest value of $P$) as
the probable identification.
Nevertheless, there are strong circumstantial arguments
for the other source being the correct identification because
it is both a
15$\mu$m ISO source (W2003)
and the galaxy associated with the source has a disturbed
morphology (see the K-band image in Figure 1).
The second radio source is, however, much closer to the
SCUBA position. 
There is nothing on either the K-band or I-band images at the
position of this radio source.

\section{Reliability of Survey}

All of the teams carrying out SCUBA surveys have recognised
that some of their sources are likely to be spurious, since
many of the sources are detected with low signal-to-noise
and the surveys are often close to the confusion limit.
In earlier papers (Eales et al. 2000; W2003) we
estimated that 10\% of the CUDSS sources are likely to
be spurious, based on both gaussian statistics and 
on the results of applying our source-detection alogorithm
to negative maps. Since any SCUBA source with a secure identification
is likely to be a genuine submillimetre source, we can
use the results of the previous sections to investigate empirically
the reliability of the CUDSS survey. In this section we also compare
the results of our identification analysis with the results
of a similar analysis for the `8 mJy' SCUBA survey, the
other large-area blank-field SCUBA survey.

The 8-mJy survey contains 36 sources detected at $>$3.5$\sigma$
with 850$\mu$m flux densities $\geq$8 mJy in an area of sky
of 260 arcmin$^2$ (Scott et al. 2002). Thus the survey is 
less sensitive but covers a
larger area than CUDSS. 
The expected percentage
of spurious sources, based on gaussian statistics, is about half
the value expected for CUDSS. Ivison et al. (2002) have shown
that the fraction of 8-mJy sources with radio associations
drops systematically in areas of the original
submillimetre images with high noise. They have used
this fact to argue that six of the 8-mJy sources are likely
to be spurious. After removing these six sources from the catalogue,
they find that 60\% of the 30 remaining 8-mJy sources are detected
at radio wavelengths.

The first thing we can do is compare the percentages of SCUBA sources
with radio associations in the two surveys. Nine out of 27 sources
in the CUDSS 3-hour field and five out of 23
sources in the CUDSS 14-hour field (Eales et al. 2000; Webb et
al. 2003a) have radio detections. These are much lower percentages
than are found for the 8-mJy survey. However, the difference can
almost certainly be attributed to the different sensitivities,
at both submillimetre and radio frequencies, of the different
surveys. To show this, we have compared the radio and submillimetre
surveys of the CUDSS 3-hour field with the corresponding
8-mJy surveys. We
have excluded the CUDSS 14-hour field because the radio observations
were made at 5 GHz, which makes the comparison difficult.
Let us assume that the redshift distributions of the
objects in the CUDSS and 8-mJy survey are similar. 
The 8-mJy sources are, on average, a factor of $\sim$2 brighter
than the CUDSS sources at 850$\mu$m. If the redshift distributions
are the same, they will also be brighter by a similar factor
at 1.4GHz. 
To investigate the effects of the flux limits, we have decreased
the radio flux of each 8-mJy source (Ivison et al. 2002) by this factor.
This decrease now makes the 8-mJy sources directly comparable to
the CUDSS.
Only seven of the 8-mJy sources now have radio fluxes
which fall above the 40$\mu$Jy limit of our radio observations.
Thus the higher percentage of radio detections
for the 8-mJy survey is entirely the result of the different
flux limits of the two surveys.

We have followed Ivison et al. (2002) in using the statistics of the
radio detections to investigate the reliability of the CUDSS survey.
Figure 3 shows the fraction of SCUBA sources that are also
radio sources as a function of both the signal-to-noise in the original
submillimetre survey and of noise in the original submillimetre
image. 
Because of the small
number of sources, we have expressed this as a cumulative fraction.
In the 3-hour field, the fraction of radio detections clearly does
not depend on signal-to-noise.
In this field the radio fraction does appear to increase at low values of the
submillimetre noise, but since this increase is due to only
four out of the 27 sources (of which two are radio detections),
we do not regard it as significant.
In the 14-hour field, the radio fraction does appear to 
depend on signal-to-noise but does not depend on the value
of the submillimetre noise.
Ivison et al. used the result that the
radio fraction falls with increasing submillimetre noise
in both of the 8-mJy fields to eliminate sources in regions
of high noise. Since there is no similar effect which occurs in
both of the CUDSS fields, we conclude there is no compelling
statistical
evidence to eliminate CUDSS sources below some signal-to-noise
threshold or in regions of high noise.

\begin{figure*}
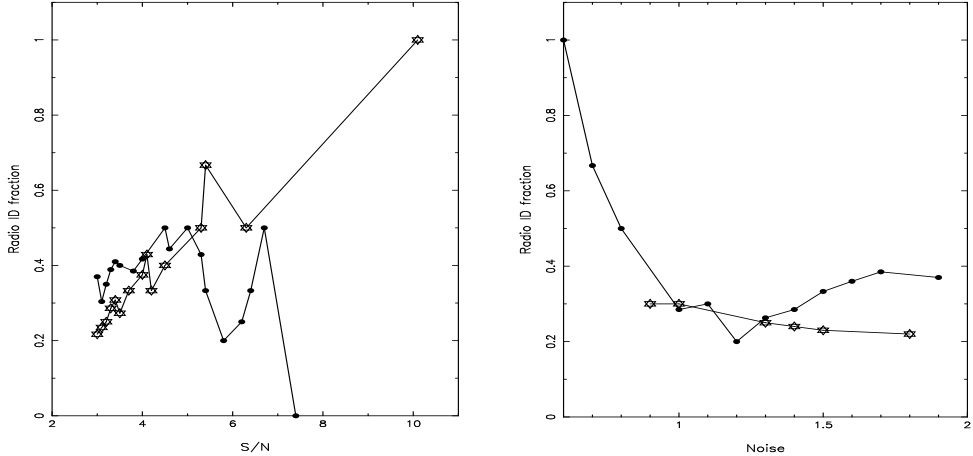


\subfigure{\psfig{file=Fig3a.ps,height=6.0cm,width=6.0cm}}
\goodgap
\subfigure{\psfig{file=Fig3b.ps,height=6.0cm,width=6.0cm}}\\

\caption{
Plots of cumulative fraction of SCUBA sources with radio detections
against the signal-to-noise with which a source was detected
in the original SCUBA survey (lefthand plot) and against the noise in
the original SCUBA survey at the position of the source (righthand plot).
The stars represent the results for the 14-hour field, the circles
the results for the 3-hour field.
}

\end{figure*}

A final useful thing we can do with the identification statistics
is to derive empirically the distribution of SCUBA position errors.
In an earlier paper (Eales et al. 2000), we used a Monte-Carlo
simulation to predict the distribution of position errors, but
it is preferable to determine these directly. Since the radio
positions have an accuracy of better than one arc second, the
offset between the
radio position and the submillimetre position of a source is
a direct measurement of the error in the submillimetre position.
There is one caveat to this. If there are errors greater than 8 arcsec,
we will miss them, because that was the maximum distance 
out to which we looked for radio sources (\S 4). Figure 4a shows
the histogram of positional errors for the 14 CUDSS sources with
radio detections, overlaid with the 
distribution of errors predicted from our Monte-Carlo
simulation (Eales et al. 2000). The figure shows that in
practice our positions are slightly more accurate than the
Monte-Carlo simulation predicted. For example, three out
of 14 sources (21\%) have positional errors $\leq$1 arcsec,
whereas our simulation predicted that there would
be essentially no sources with position errors this
small.

Figure 4b shows the positional errors derived in the same
way for the 8-mJy sample (Ivison et al. 2002). 
In the case of the 8-mJy sources, we may be slightly biased
towards small positional errors, because the deeper radio data
means that a source $>$4 arcsec from the submillimetre position
can not always be confidently associated with the submillimetre
source (Ivison et al.).
Since the 8-mJy survey is further from the submillimetre
confusion limit than the CUDSS, one might expect the accuracy of the
positions to be rather better. 
However, the CUDSS positions
are at least as good. For example, five out of 14 CUDSS sources
have positional errors $\leq$2 arcsec compared with four out of 18
8-mJy sources. This comparison lends some support to the elaborate,
if not very elegant, cleaning
technique we used to produce the source catalogue (Eales et al. 2000; W2003).

\begin{figure*}
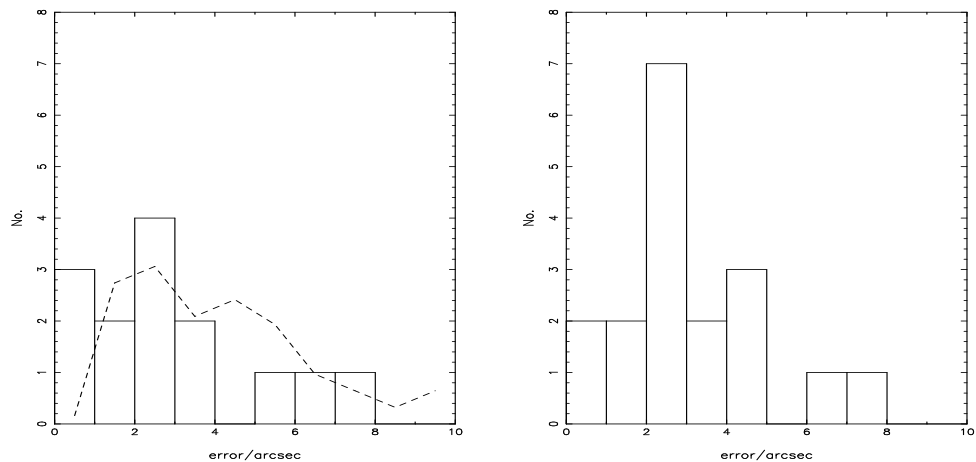


\subfigure{\psfig{file=Fig4a.ps,height=6.0cm,width=6.0cm}}
\goodgap
\subfigure{\psfig{file=Fig4b.ps,height=6.0cm,width=6.0cm}}\\

\caption{
Offsets between the radio and submillimetre positions for
submillimetre sources which have radio detections. 
Since the radio positions are very accurate, these offsets
are effectively the errors on the submillimetre positions.
The lefthand
figure is for the CUDSS. The dashed line shows the prediction
we made for the CUDSS positional errors from a Monte-Carlo
simulation (Eales et al. 2000). The righthand figure shows the
same histogram for the 8-mJy sample, using the data in Ivison
et al. (2002).
}

\end{figure*}

\section{The identifications at low redshift---gravitational lensing?}

Our identification analysis is based entirely on calculating the
probability that an object---either a radio source or a faint galaxy---would
fall so close to the position of a SCUBA source if it
were actually physically unrelated to
the SCUBA source. There
is one possible weakness in this approach. If a significant fraction
of the SCUBA sources are gravitationally lensed, then it is
possible that this technique will find the lens rather than the
galaxy which is genuinely responsible for the submillimetre emission.
Since the lens will always be at a lower redshift, this method could
produce spurious low-redshift identifications for SCUBA sources
(Chapman et al. 2002). 

The CUDSS contains a larger number 
of low-redshift identifications than were found in
the 8-mJy survey (Ivison et al. 2002). Our small pilot survey at
an RA of 10 hours contained two sources which have identifications
with spectroscopic redshifts $< 1$ ($z=0.074$ and $z=0.55$,
Lilly et al. 1999).
The two large fields contain three sources with identifications
with spectroscopic
redshifts below this limit ($=0.088$, $z=0.176$, $z=0.66$)
plus two sources with identifications with estimated redshifts 
(from the broad-band colours---\S 9.2) below this limit.
Might some of these objects actually
be a lens rather than being the galaxy responsible for the
submillimetre emission?

There are three arguments that this is not generally the
case. The first of these is described in detail in \S 9, in
which we show that the redshift estimated for the galaxy
from multi-band photometry is generally very similar to the
redshift estimated for the SMS from the ratio of radio-to-submillimetre
flux. The second argument is based on the morphologies of
the galaxies. 
If the galaxies are lenses, they should be galaxies 
which just happen to fall between the SCUBA source and the Earth,
with the only bias in their properties being that
they will tend to be galaxies which produce large
gravitational amplification factors.
There is no reason to expect them to have the morphological peculiarities
characteristic of ULIRGs or SCUBA galaxies. However, many of our low-redshift
galaxies are indeed extremely peculiar systems. Good examples are the 
systems of interacting galaxies CUDSS 3.8 and 3.10 
(Figure 2). 

The third argument is based on the fact that most of the low-redshift
identifications are also ISO 15$\mu$m sources.
We have 15$\mu$m observations of the 3-hour
and 14-hour fields but not of the 10-hour field. Of the five
low-redshift objects
in the former fields,
four are detected at 15$\mu$m
(Note that we did not use the 15$\mu$m results in
our identification analysis---\S4).
The typical shape of the spectral energy distribution
of galaxies 
means that galaxies are unlikely to be detected at 15$\mu$m at
$\rm z \geq 1$ (Eales et al. 2000; Flores
et al. 1999). It is therefore unlikely that a 
SCUBA source at $\rm z >>1$ 
which is being lensed by a low-redshift galaxy will be
detected at 15$\mu$m.
It is possible that the SMS is being lensed but the 15$\mu$m emission is from
the lens rather than the SMS. 
By comparing the surface density of 15$\mu$m sources with the
surface density of galaxies with $\rm K < 20$
(Cowie et al. 1994; Flores et al. 1999), we estimate
that the probability of a lens also being
a 15$\mu$m source is roughly 10\%. 
Therefore, the probability that four out of five lenses are also 15$\mu$m
sources is clearly extremely low.

One source, CUDSS 3.2, may be the exception that proves the rule.
It is the one low-redshift SMS which is not detected at 15$\mu$m;
the optical counterpart has an undisturbed morphology; and
the redshift
estimated for the counterpart
from multi-band photometry (0.62,\S 9.2) is much lower than
the redshift limit estimated from the ratio of 
radio-to-submillimetre flux ($>1.7$, \S 9.1).
These properties are all consistent with the hypothesis
that the optical counterpart is actually a lens which is gravitationally
amplifying the radio and submillimetre emission from an SMS at a
much higher redshift. However, the difference in all three respects between
this object and the other low-redshift counterparts strongly
suggests that the latter SMSs are genuinely at low redshift.

Of the 50 SMSs in the two large fields, there is only one source
for which there is plausible evidence for lensing. Blain (1998)
predicted that about 2\% of SMSs with $\rm S_{850 \mu m} \simeq 10 mJy$
are gravitationally amplified by a factor of $\geq$2. This is
the same as our observed fraction, although the prediction is
for a different flux level.

We cannot not use the arguments above
for the two sources in the 10-hour field because we have no
15$\mu$m data for this field. However, both these sources are
also detected at 450$\mu$m. Since the ratio of 450 and 850$\mu$m
flux is expected to fall with redshift (Figure 8 of Eales et al. 2000),
the detection of these sources at 450$\mu$m is strong circumstantial
evidence that the low-redshift identifications are correct.

If the gravitational-lensing hypothesis can be eliminated, is there
any other effect which might produce spurious low-redshift
identifications? There is one effect which might be important.
At the submillimetre flux level of the CUDSS, the confusion of faint
sources is likely to be important. Our Monte-Carlo simulations
(Eales et al. 2000) revealed the possible importance of `flux-boosting',
in which an apparent single source is actually two or more sources,
which are only in the survey because their combined fluxes 
are greater than the flux limit of the survey. If this is the case for any
of our sources, then it is possible that there are two or more 
genuine identifications close to the submillimetre position, but
we have only found the identification at the lower redshift.
There is one possible example of this.
CUDSS 3.27 has two possible counterparts, each with a low value
of $P$ (\S 5). In this case, we rejected the counterpart which is detected
at 15$\mu$m because it has a higher value of $P$ than the alternative.
But it is possible that both identifications are correct.

Finally, we note that although the fraction of sources with low-redshift
identifications is higher in the CUDSS than in the 8-mJy survey, the estimated
redshifts of the sources in the 8-mJy survey with submillimetre fluxes below
8mJy is, on average, 0.6 lower than the estimated redshifts of the
sources above this flux limit (Ivison et al. 2002). This is additional evidence
that the phenomenom that the fraction of sources with low redshifts is increasing
as the submillimetre flux limit decreases is a genuine one.

\section{The Nature of the Identifications---Morphologies and Colours}

There are 14 secure identifications in the 3-hour and 14-hour fields
(this paper and Webb et al. 2003a). Of these, one is not detected at infrared
or optical wavelengths, and so it is impossible to classify the
morphology of the galaxy; two show no signs of an interaction or
have no morphological peculiarity; the remaining 11 show some
signs of an interaction or have some peculiarity in the structure.
Ivison et al. (2002) performed a similar analysis for the 8-mJy
sample. Of the 21 secure identifications, they listed six as being
too faint at optical/infrared wavelengths to classify
morphologically; 13 as being distorted
or close multiple systems; and two as being compact. Given the subjectivity
in making classifications of this kind, the proportions seem quite similar
in the two surveys.

We are on stronger ground in classifying galaxies according
to their colours. Ivison et al. (2002) divided galaxies into
EROs ($I-K > 4$) and VROs ($3.3 < I-K < 4.0$). Of 18 SCUBA sources
with radio detections, they found that seven objects had normal colours,
10 objects could be classified as either a VRO or an ERO, and one
source was not detected at optical/infrared wavelengths.
Of our 14 secure identifications, we find five objects with normal
colours, eight objects which are either VROs or EROs, and one object
which is not detected at infrared or optical wavelengths.
The proportions of objects in the different classes are thus remarkably
similar for the two samples.

\section{Estimating Redshifts}

Chapman et al. (2003a) have recently succeeded in using the
Keck Telescope to measure 
redshifts for 10 SCUBA sources, the first significant
number of SCUBA galaxies for which this has been done.
However, despite this important success, it is likely that methods
for estimating redshifts will be important for several
years to come. First, Chapman et al. only targeted SCUBA
sources which were detected at radio wavelengths and
had I magnitudes $\rm 22.2 < I < 26.4$, and thus their results are
strictly applicable only to the $\sim$50\% of the SCUBA population
that satisfy these limits. Since the ratio of radio-to-submillimetre
flux is expected to fall with redshift (Carilli and Yun 1999), the
radio
criterion, in particular, is
likely to lead to an underestimate of 
the proportion of SCUBA sources with
$\rm z \geq 2$. Second,
Chapman et al. only succeeded in measuring redshifts for about 30\% of the
sources which satisfied the above criteria.
The sources for which they failed may either have weak emission lines
or be at a redshift at which emission lines are hard to detect
(Chapman
et al. noted
the relative lack of SCUBA galaxies in the redshift
range $\rm 1 < z < 2$, the so-called `redshift desert', a redshift
interval in which few strong emission lines fall in the optical
waveband). For these reasons, methods
for estimating redshifts of SCUBA galaxies are 
likely to continue to be important.

In this section, we investigate two methods for estimating redshifts.
Both are well-known but only one has been applied before to
SCUBA galaxies. In both cases, we have used the spectroscopic redshifts
that do exist for SCUBA galaxies, both from the work of Chapman et
al. and from our own work, to test the efficiency of the methods.

\subsection{The Radio Method}

Carilli and Yun (1999) were the first to point out that for a
star-forming galaxy the ratio of radio-to-submillimetre
flux should be a function of redshift, and thus that it
should be possible to estimate the redshift of a star-forming
galaxy from this ratio.
Following the original suggestion, a
number of groups
used different samples of low-redshift objects to
determine the expected relationship between
this ratio
and redshift
(Carilli and Yun 2000; Dunne, Clements and
Eales 2000;
Rengarajan and Takeuchi 2001).
There are slight differences between the redshifts estimated
using the different sets of low-redshift templates
(Ivison et al. 2002).

Figure 5
shows the ratio of submillimetre to radio flux plotted against
redshift for all SCUBA galaxies which have
both spectroscopic redshifts and
radio measurements. We have plotted on the figure the predictions
for star-forming galaxies using the 104 low-redshift templates
of Dunne, Clements, and Eales (2000). As described in that paper,
we first predict the relationship between the flux ratio and
redshift for each template and then, at each redshift, determine
the median and $\pm$1$\sigma$ predicted values. The one slight
difference from that paper is that the templates have been modified
to incorporate our 450$\mu$m observations of the galaxies (Dunne
and Eales 2001).

\begin{figure}
\psfig{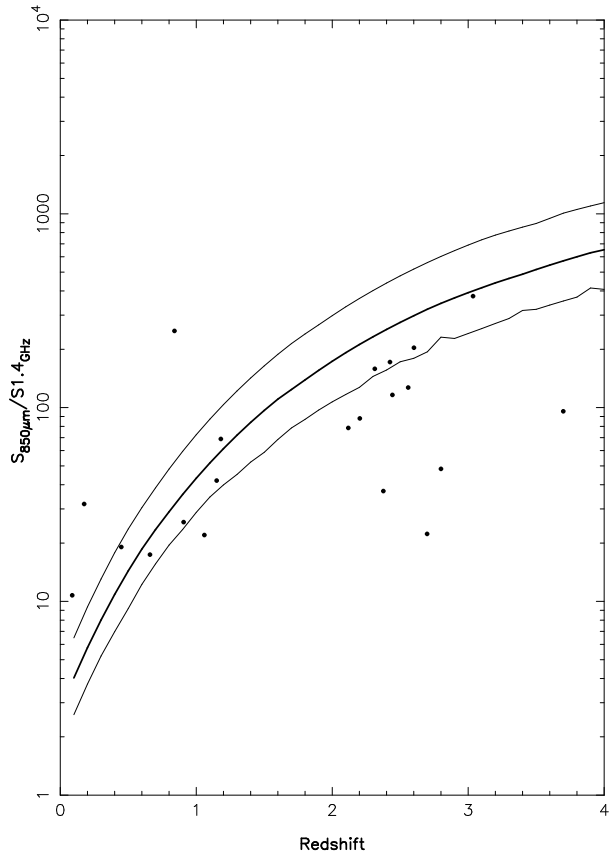}
\caption{The ratio of
850$\mu$m flux to 1.4-GHz flux verses redshift.
The lines show predictions of how this flux ratio
should depend on redshift for star-forming galaxies
using the method described in Dunne, Clements and Eales
(2000) and in the text.
The thick line shows the median prediction of the
templates and the
thin lines show $\pm1\sigma$ predictions based on the range
of predicted values at each redshift.
The points show SMSs with spectroscopic redshifts and radio detections.
The data are from Eales et al. (2000),
Smail et al. (2000), Ivison et al. (2002), Chapman et al. (2002,
2003), Simpson et al.
(2003) and this paper.
}
\end{figure}

At first sight, the diagram does not 
instill one with much confidence in the method, since for the high-redshift
data there is not even  a strong correlation between the
the flux ratio and redshift. The diagram also suggests
that redshifts estimated in this way will generally
be underestimates, since nine sources lie below the
$\pm 1\sigma$ predictions, while only three sources lie
above these predictions.
Figure 6 shows
the difference between the spectroscopic redshift and
the redshift estimated from the median redshift in
Figure 5. This figure shows that for about half the SMSs
the method works quite well, leading to redshift
errors of $\rm z < 0.5$. However, there are also a significant
number of SMSs where the method results in a catastrophic
redshift error. 
Similar results are obtained if the other sets of low-redshift
templates are used.
In Table 3 we have listed redshifts estimated
in this way for the 15 CUDSS sources 
with secure identifications.

\begin{figure}
\psfig{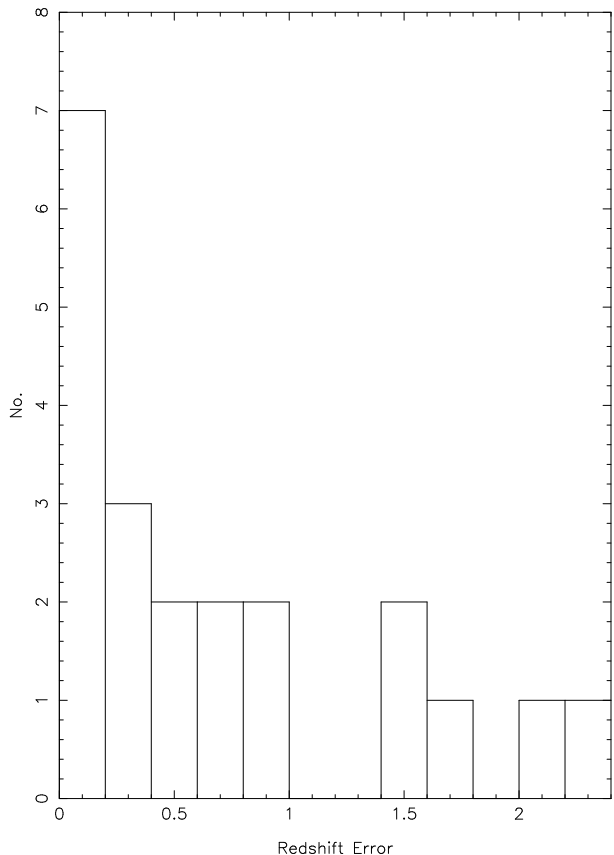}
\caption{The difference between the spectroscopic redshift
and the redshift estimated from the radio method for the
21 SMSs with spectroscopic redshifts and radio measurements.
We have estimated the redshift of each SMS using the median
prediction in Figure 5.
}
\end{figure}

\subsection{Photometric Redshifts}

A method which has not been used before to estimate the
redshifts for SCUBA galaxies is to use multi-band optical
and infrared photometry to estimate `photometric redshifts',
a method which was first used extensively in the studies
of the Hubble Deep
Field. We are in a good position to examine the utility of
this method for SCUBA galaxies, because we have observations
in five different photometric bands: U, B, V and I photometry
from the Canada France Deep Fields and our own K-band photometry.
Table 3 lists the multi-band photometry for our 14 secure identifications
and multi-band photometry for two other SCUBA galaxies which
have spectroscopic redshifts.

We have used the photometric redshift programme of Benitez (2000).
Starting from an ensemble of SEDs for low-redshift galaxies,
the programme determines the redshift and SED which provide
the best fit to the multi-band photometry of the galaxy in
question. The attractive feature of the programme is that it
uses Bayes theorem to incorporate some prior knowledge about
the galaxy population, an approach which reduces the number
of `catastrophic' redshift errors. The programme does not
incorporate any theoretical assumptions about galaxy evolution and
does not allow for the possibility of dust reddening. However,
it is impressively successful at matching the spectroscopic redshifts
in the Hubble Deep Field (Benitez 2000) and in the Canada-France
Redshift Survey (Waskett et al. 2003b). Table 3 lists the redshifts
estimated using this programme.

SCUBA galaxies, of course, must contain large amounts of dust,
and thus one might expect any photometric redshift technique
to break down when dealing with objects like this. Figure 7 shows
the redshift estimates plotted against the spectroscopic redshifts
for the six SCUBA galaxies with both spectroscopic redshifts
and enough multi-band data to make the photometric technique
worth while. The error bars on the photometric redshifts show
the redshift range in which there is a probability of 95\% 
that the true redshift lies. For five of the six sources, the
agreement between the spectroscopic and photometric redshift is
very good, and for the remaining source
the disagreement is within the range of the errors.
Therefore, although this is a small sample, we conclude that
estimating the redshifts of SCUBA galaxies from multi-band
photometry is at least as accurate as estimating the redshifts
from the ratio of radio-to-submillimetre flux.

Figure 8 shows the two sets of redshift estimates plotted against each
other. With the exception of CUDSS 3.2, there is 
surprisingly good agreement between the two sets,
suggesting that for the CUDSS sources we can have some confidence
in our redshift estimates. We note that many of the estimated
redshifts lie in the so-called `redshift desert', $\rm 1 \leq z \leq 2$,
a range for which there are no bright emission lines
in the optical waveband. It may therefore be quite difficult to measure
redshifts for some of these galaxies.

The good agreement between the two sets of redshift estimates is the
third piece of evidence that gravitational lensing is not
generally important (\S 7). If lensing were important, there would be
no reason why the estimates should agree, since the photometric-redshift
method would yield the redshift of the lens and radio-to-submillimetre
method would yield the redshift of the lensed object.

\begin{table*}
\begin{tabular}{|l|l|l|l|l|l|l|l|l|l}
\multicolumn{10}{|c|}{Table 3. Magnitudes and Redshifts}\\
\hline
(1)&(2)&(3)& (4) & (5) & (6) & (7) & (8) & (9) & (10) \\
Name & $U_{AB}$ & $B_{AB}$ & $V_{AB}$ & $R_{AB}$ & $I_{AB}$ & $K_{AB}$ & $\rm z_{phot}$ 
& $\rm z_{spec}$ & $\rm z_{radio}$ \\
\hline

3.2 & $\rm 23.63\pm0.05$ &  $\rm 23.33\pm0.02$ & $\rm 22.63\pm0.02$ & ...& 
$\rm 21.38\pm0.01$ & 
$\rm 20.55\pm0.02$ & $\rm 0.62\pm0.21$ & ... & $\rm >1.7$ \\
3.6 & $\rm >26.98$ &  $\rm >26.38$ & $\rm 26.48\pm0.41$ & ... & 
$\rm >25.62$ &
$\rm 23.39\pm0.21$ &
$\rm 1.57_{-0.46}^{+0.76}$ & ... & $\rm 1.35\pm0.33$ \\
3.7 & $\rm >26.98$ & $\rm >26.38$ & $\rm >26.40$ & ... &
$\rm >26.52$ & $\rm 22.36\pm0.09$ & ... & ... & $\rm 2.1\pm0.6$ \\ 
3.8 & $\rm 20.09\pm0.01$ & $\rm 19.09\pm0.003 $ & $\rm 18.25\pm0.002$ &
... & $\rm 17.33\pm0.001$ & $\rm 16.47\pm0.003$ & $\rm 0.25\pm0.16$ & 0.088 & 
$\rm 0.4\pm0.18$ \\
3.10 & $\rm 21.20\pm0.01$ & $\rm 20.54\pm0.01$ & $\rm 19.07\pm0.004$ & 
... & $\rm 19.19\pm0.002$ & $\rm 18.22\pm0.005$ & $\rm 0.40\pm0.18$ & 0.176 &
$\rm 0.85\pm0.23$ \\
 & & & & & & & & \\
3.15 & $\rm 24.39\pm0.07$ & $\rm 23.75\pm0.02$ & $\rm 23.31\pm0.03$ &
... & $\rm 21.90\pm0.01$ & $\rm 20.24\pm0.02$ & $\rm 0.73\pm0.23$ & ...
& $\rm 0.6\pm0.18$ \\
3.17 & $\rm >26.98$ &  $\rm >26.38$ & $\rm 26.48\pm0.41$ & 
... & $\rm >25.62$ & $\rm >22.84$ & ... & ... & $\rm 1.60\pm0.42$ \\ 
3.24 & $\rm 25.63\pm0.16$ & $\rm 24.85\pm0.06$ & $\rm 24.73\pm0.08$ &
... & $23.22\pm0.02$ & $\rm 21.17\pm0.07$ & $\rm 1.14\pm0.28$ & 
... & $\rm 1.0\pm0.25$
\\
3.25 & $\rm 25.29\pm0.14$ & $\rm 25.62\pm0.21$ & $\rm 24.72\pm0.08 $ 
 & ... &  $\rm 23.82\pm0.03$ & $\rm 22.47\pm0.17$ & 
$\rm 1.05_{-0.65}^{+0.27}$ & ... & $\rm 0.4\pm0.18$ \\
14.1 & $\rm 27.17\pm0.32$ & $\rm 26.60\pm0.12$ & $\rm 26.28\pm0.11$ &
... & $\rm 24.71\pm0.04$ & $\rm 21.18\pm0.03$ & $\rm 1.25\pm0.3$ & ... &
$\rm 1.9\pm0.48$ \\

 & & & & & & &  & & \\

14.3 & $\rm 24.71\pm0.06$ & $\rm 24.55\pm0.03$ & $\rm 24.06\pm0.02$ &
... & $\rm 23.19\pm0.01$ & $\rm 21.23\pm0.04$ & $\rm 1.11\pm0.28$ & ... &
$\rm 1.11\pm0.28$ \\
14.9 & $\rm >26.98$ & $\rm 26.62\pm0.12$ & $\rm 26.40\pm0.13$
& ... &  $\rm 24.89\pm0.05$ & $\rm 21.12\pm0.03$ & $\rm 1.44\pm0.32$ & ... & 
$\rm 1.7\pm0.43 $ \\
14.13 & $\rm 23.93\pm0.05$ & $\rm 23.73\pm0.02$ & $\rm 22.90\pm0.01$ &
... & $\rm 20.86\pm0.004$ & $\rm 18.42\pm0.03$ & $\rm 0.90\pm0.25$ & 1.15 & 
$\rm 1.2\pm0.3$ \\ 
14.18 & $\rm 22.97\pm0.03$ & $\rm 22.57\pm0.01$ & $21.99\pm0.01$ &
... & $\rm 20.61\pm0.003$ & $\rm 18.95\pm0.01$ & $\rm 0.69\pm0.22$ & 0.66 & 
$\rm 0.7\pm0.2$ \\

 & & & & & & &  & & \\
N2 850.4$^a$ & ... & ... & $\rm 22.40\pm0.03$ & $\rm 22.47\pm0.01$ & 
$\rm 22.45\pm0.02$ & $\rm 18.43\pm0.02$ & $\rm 1.3_{-0.3}^{+1.18}$ &
2.376 & ... \\
N2 850.8$^a$ & ... & ... & $\rm 22.79\pm0.03$ & $\rm 22.68\pm0.02$ &
$\rm 22.17\pm0.02$ & $\rm 20.06\pm0.09$ & $\rm 1.41\pm0.32$ & 1.189 & ... \\

\hline
\end{tabular}
\flushleft
(1) Source name. (2)-(7) Magnitudes in the AB system in the different
photometric bands. Except where noted,
the optical magnitudes are from the Canada-France
Deep Fields survey (\S 3) and the infrared magnitudes are from this
paper. In both cases, the errors on the magnitudes
do not include the calibration error, which is about
0.05 mags. (8) Redshift estimated using the photometric redshift
method of Benitez (2000). (9) Spectroscopic redshift. 
(10) Redshift estimated
from the ratio of radio-to-submillimetre flux. 

Notes on sources: a---The data for
these objects were taken from Ivison et al. (2002).
\end{table*}

\begin{figure}
\psfig{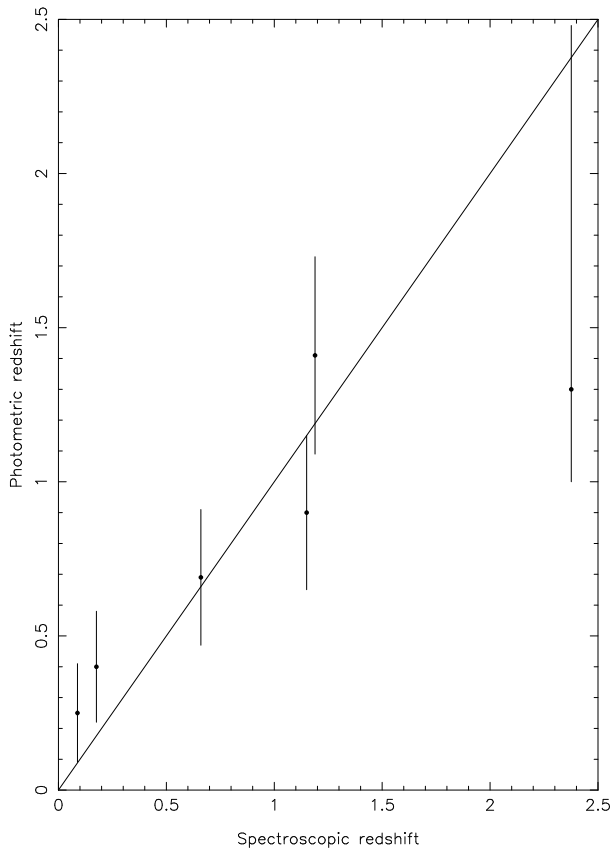}
\caption{Spectroscopic redshift verses photometric redshift
for the six galaxies with extensive multi-band photometry and
spectroscopic redshifts. The spectroscopic redshift is equal to
the photometric redshift along the line.
}
\end{figure}

\begin{figure}
\psfig{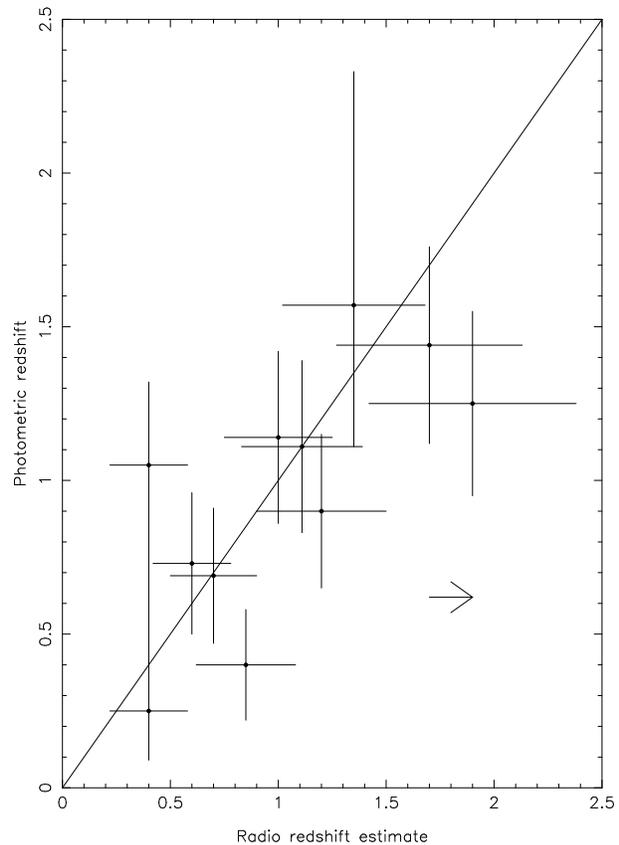}
\caption{Redshift estimated from the multi-band optical and
infrared photometry verses redshift estimated 
from the radio-to-submillimetre flux ratio. The arrow
is CUDSS 3.2, which has an upper limit for its
radio flux.
}
\end{figure}

\section{Discussion}

In this section we will discuss what the optical, infrared and
radio observations of the CUDSS sources reveal about the nature
of SMSs. A later paper will describe an investigation of the
evolution of the submillimetre luminosity function which
will incorporate the new results.

A simple thing we can do is compare the 
far-IR---submillimetre luminosities of
the CUDSS sources with the 
luminosities of dust sources in the local universe.
A problem which is often skated over in calculating the
luminosity of SMSs is that there
is usually a flux measurement at only a single wavelength,
and therefore the calculation of the luminosity requires some
assumption about the SED of the SMS. To investigate the
effect of this assumption, we have calculated the luminosity
of the CUDSS sources making two different assumptions about the
SED. We used two
extreme
SEDs from the sample of IRAS galaxies of Dunne et al.
(2000). NGC 958 is a galaxy whose SED
is dominated by cold dust.
The observed fluxes of this galaxy are fitted well
by the two-component dust model of Dunne and Eales (2001), with dust
at 20K and 44K in the ratio by mass of 186:1. At the other
extreme is the galaxy IR1525$+$36, which, in the Dunne and Eales model, has
dust at 19K and 45K in the ratio by mass of 15:1. Figure 9 shows
the luminosities of the CUDSS sources with secure identifications
calculated using these two different assumptions. We have
also plotted in the figure the luminosities
of the IRAS galaxies in the sample of Dunne et al. (2000). 
For the CUDSS galaxies without spectroscopic
redshifts, we have used the redshift estimated
from our multi-band photometry (\S 9.2) and, if that is not possible,
the redshift estimated from the ratio of radio and submillimetre
flux (\S 9.1). The figure shows that there is roughly a factor of
five difference in the luminosities of the CUDSS sources calculated
with the two different SEDs, showing the sensitivity of
the calculation to the assumption about the SED. With the cold SED,
there is a substantial overlap in the luminosities of the CUDSS
sources with the low-redshift sample, although the majority
of the CUDSS sources are still more luminous than the most luminous
object in the local sample, the archetypical ULIRG Arp 220.

\begin{figure}
\psfig{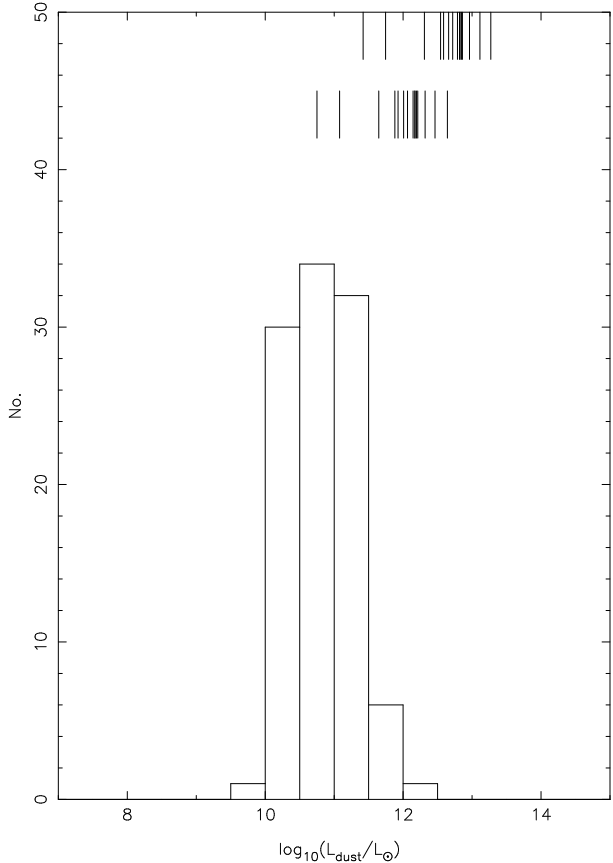}
\caption{The far-IR---submillimetre
luminosities of the CUDSS sources with secure identifications
and of the IRAS galaxies in the sample of Dunne et al. (2000).
The histogram shows the luminosities of the IRAS galaxies.
The lower set of vertical lines mark the luminosities of the
CUDSS sources calculated using the SED of NGC 958 (see text);
the upper set of lines show the luminosities calculated
using the SED of IR1525$+$36.
}
\end{figure}

Given our extensive multi-band optical/IR photometry, 
we can calculate the ratio of dust luminosity to
optical/IR luminosity.
For each source with a secure identification,
we calculated the optical/IR flux by integrating the observed
SED from 0.25 to 2.5$\mu$m. We estimated the flux at each wavelength
by making a power-law interpolation between the two
neighbouring photometric measurements. The biggest uncertainty
in this calculation is the question of which SED to use
to calculate the dust luminosity. Figure 10 shows
the histogram of dust luminosity divided by optical/IR luminosity,
with the dust luminosity calculated using the cold SED.
The figure shows that most of the CUDSS sources have dust luminosities
which are between 10 and 100 times greater than emission in the
optical/near-IR bands. If the hot SED is used to calculate the
dust luminosities, these figures increase by a factor of about five.
Whichever SED is used, figures 9 and 10 show that, as one would expect, 
the CUDSS sources are luminous systems with most of the emission being
reprocessed emission from dust.

\begin{figure}
\psfig{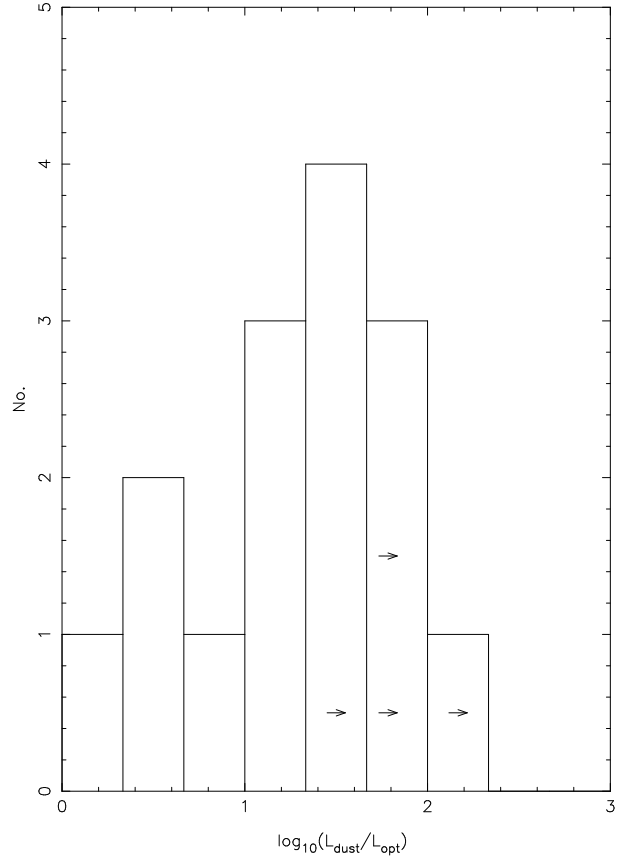}
\caption{Histogram of the ratio of dust luminosity to
optical/IR luminosity for the 
CUDSS sources with secure identifications.
We have calculated the dust luminosity using the SED of NGC 958
(see text). The lower limits are for sources which have
not yet been detected in the optical/IR.
}
\end{figure}

We now compare the absolute magnitudes
of the SMSs with those of other high-redshift objects. Dunlop (2002)
plotted the K magnitudes of SMSs against their redshift
and compared this diagram to the same diagram for radio galaxies,
which are among the most luminous galaxies
known. By comparing the apparent magnitudes of SMSs and radio
galaxies at the same redshift, he was able to compare the
absolute magnitudes of the two types of object. 
He concluded that the host galaxies of SMSs
have absolute magnitudes which are very similar to those of
radio galaxies.
At redshifts $<3$, the K-band falls on the long-wavelength
side of the 4000\AA\ break, and thus the K-band light is not
dominated by the light from young stars but rather by the light
from the stars that form most of the stellar mass of
a galaxy. Therefore, one
inference which one might draw from Dunlop's result is that the
host galaxies of SMSs are giant galaxies in which a large
fraction of the stars have already formed.

A limitation of this study, however, was that at that time
there were only three SMSs with spectroscopic redshifts. These were
also SMSs which were
known to be gravitationally
lensed,
which means there is necessarily some uncertainty
in the value of the gravitational amplification factor. Because
there are now a signficant number of SMSs with spectroscopic
redshifts, we can now carry out a much more extensive comparison
of the magnitudes of SMSs with the magnitudes of other high-redshift
objects. 

Figure 11 shows K magnitude plotted against redshift for
(a) all SMSs with spectroscopic redshifts which are not
known to be lensed and (b) all CUDSS sources with secure
identifications. For the CUDSS galaxies without spectroscopic
redshifts, we have used the redshift estimated
from our multi-band photometry (\S 9.2) and, if that is not possible,
the redshift estimated from the ratio of radio and submillimetre
flux (\S 9.1). We have also plotted on the diagram the data for
radio galaxies described in Eales et al. (1997). In order to
ensure that there are no spurious differences caused by magnitudes
being measured in apertures of different sizes, we have corrected
all the magnitudes to a common metric aperture. Most of the magnitudes
for the SCUBA galaxies have been measured through a 3-arcsec aperture, 
which at $z=2$ is 
equivalent, with our cosmological assumptions (\S 1), to
a physical distance of 23.5 kpc. We have converted the photometry
for the radio galaxies to this metric aperture using the
method described in Eales et al. (1997).
The figure confirms Dunlop's conclusion that many SMSs have host galaxies
which are as luminous as radio galaxies. About half the SMSs are, however,
in host galaxies which are fainter than radio galaxies, although the
difference
is usually small enough that they must still be fairly luminous systems.

\begin{figure}
\psfig{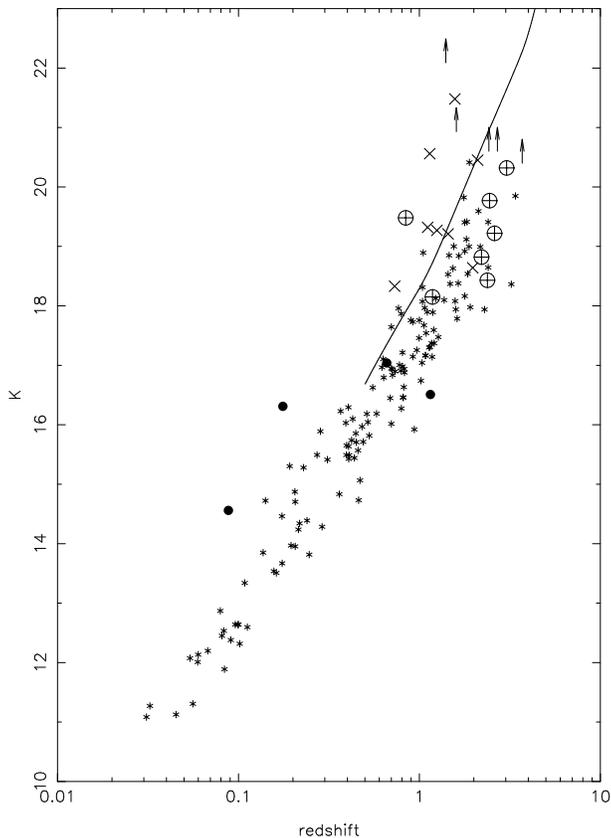}
\caption{K-band magnitude verses redshift for SCUBA galaxies
and radio galaxies. The stars shows the positions of the
samples of radio galaxies described in Eales et al. (1997).
The other symbols represent SMSs. The filled circles are for
CUDSS sources with secure identifications and
spectroscopic redshifts. The crosses-in-circles show other
SMSs with spectroscopic redshifts (Chapman et al. 2003a). The 
crosses represent
CUDSS sources with secure identification but only estimated
redshifts (see text). The lower limits are for SMSs with secure
identifications with radio sources but which are undetected
in the K band. Three of these have spectroscopic redshifts
(Chapman et al. 2003a), and two of these are CUDSS sources
for which we have estimated redshifts from the ratio of
radio and submillimetre flux. The line shows the predicted relation
for $\mu$Jy radio sources (see text).
}
\end{figure}

Another interesting population with which to compare the
SMSs are the galaxies found in $\mu$Jy radio surveys.
The morphogies of these radio sources (Muxlow et al.
2003) suggest the emission
is generally from a star-forming disk rather than being the
result of an active nucleus, as is the case for the classical
radio galaxies plotted in the figure.
Chapman et al. (2003b, and references
therein) have carried out a multi-wavelength
study of these sources.
A significant fraction of them
are also detected in the submillimetre waveband, and Chapman et al.
argue that there is a substantial overlap between the
$\mu$Jy population and the SMSs. 
They have also found the interesting result that the
optical absolute magnitudes of the $\mu$Jy radio sources have a small
range, with most of host galaxies having an optical luminosity fairly
close to $\rm L_*$. They speculate that the reason for this
may be that less luminous, and therefore less massive, galaxies
are less efficient at confining cosmic rays.

We have taken the median I-band
absolute magnitude and $I-K$ colour given in Chapman
et al. to estimate the median K-band absolute magnitude, which
we have then used to predict a $K-z$ relationship for these objects.
This is shown in the figure. It passes neatly through the middle of the
SMS points, which suggests that the host galaxies of SMSs and
$\mu$Jy radio sources are very similar in their optical/IR
luminosities. The colours of the two classes are also quite
similar. The median $I-K$ colour of the CUDSS sources with secure
identifications is 3.3, very similar to the value of 3.4 given
by Chapman et al. for the optically-faint $\mu$Jy radio sources.
Therefore, both the colours and the optical/IR luminosities
of the host galaxies are additional pieces of evidence that there
is, at the least, a substantial overlap between the two populations.

It might be thought that the identification of SMSs with luminous galaxies
is an argument against these objects being at an early stage of
galactic evolution, since a large number of stars have clearly already
formed. This is not necessarily so. Simple models of the evolution
of dust in a galaxy (Dunne, Eales and Edmunds 2003, and references
therein) imply that the mass of dust in a galaxy will be at a maximum
when roughly half the stars are formed. With the caveat
that the submillimetre
luminosity also depends on dust temperature, the time when the dust
mass is at its greatest will also be the time at which the
submillimete luminosity is at its peak. How close this time
is to the time at which star formation started in the galaxy depends
on the characteristic timescale of star formation. If most
of the stars form in a burst, as may well be the case for
elliptical galaxies, the interval between the onset of star
formation and time when half the stars have formed may be
very short indeed. If these ideas are correct, then the SMSs plotted
in the figure will be roughly a factor of two more luminous in the
optical/near-IR waveband by the current epoch.

\section{Conclusions}

We have presented optical, near-infrared and radio observations
of the 3-hour field of the Canada-UK Deep Submillimetre Survey.
We have reached the following conclusions:

\smallskip

\noindent (1) Of the 27 submillimetre sources in this field,
9 have secure identifications with either a radio source 
or a near-IR source.
Of the 50 submillimetre sources in the two CUDSS fields,
14 now have secure identifications.

\smallskip

\noindent (2) The percentage of sources with secure identifications
is consistent with that found for the bright 8-mJy submillimetre
survey, once allowance is made for the different submillimetre
and radio flux limits.

\smallskip

\noindent (3) Of the 14 secure identifications, eight are VROs or EROs,
five have colours typical of normal galaxies, and one is a radio source
which has not yet been detected at optical/IR wavelengths.
These proportions are very similar
to those found for the 8-mJy survey.
Eleven of the identifications have optical/near-IR
structures which are either disturbed or have some peculiarity which
suggests that the host galaxy is part of an interacting system, a
similar percentage to that found for the
8-mJy survey. 

\smallskip
\noindent (4) We have examined
the reliability of
the CUDSS catalogue. 
In constrast to the result of a similar analysis for the 8-mJy survey,
we find no clear evidence that CUDSS sources with low S/N or at positions
in the submillimetre maps where the noise is high are any less
reliable than the rest of the sources. 

\smallskip

\noindent (5) We have critically examined different methods of estimating
the redshifts of SMSs. 
We show that the method of estimating
redshifts from the ratio of radio and submillimetre flux (Carilli
and Yun 1999) works
well for about 50\% of SMSs, but there are a significant number
of catastrophic errors. We show the method of estimating redshifts
from the multi-band optical and near-IR photometry works surprisingly
well.

\smallskip
\noindent (6) We conclude that the low-redshift identifications are
genuine low-redshift submillimetre sources rather than being gravitational
lenses. This conclusion is based on (i) the morphologies of the identifications,
(ii) the good agreement between the photometric redshfits of the
galaxies and the redshifts estimated from the ratio of radio to submillimetre
flux, (iii) the fact that the majority of the low-redshift identifications
are also 15$\mu$m sources.

\smallskip

\noindent (7) 
We show that many SMSs are in host galaxies which
are as bright in the near-IR as radio galaxies, which are among the
most luminous galaxies in the universe. However, on average, the host
galaxies of SMSs are slightly less bright in the near-IR than the
classical radio galaxies. They are, however,
very similar, in both their absolute near-IR/optical magnitudes and
colours, to the host galaxies of the radio sources detected in
$\mu$Jy radio surveys.

\section*{Acknowledgments}

Research by DC, SAE, RI, WKG is supported by the Particle Physics and
Astronomy Research Council. SAE thanks the Leverhulme Trust for
the award of a research fellowship during a critical phase in this
research. LD is supported by a PPARC fellowship and KW by
a PPARC studentship.

{}
\end{document}